\documentclass[12pt]{amsart}
\usepackage{amsmath,amssymb}

\newcommand{\RR}{\mathbb{R}}
\newcommand{\LL}{\mathcal{L}}

\begin{document}

\title{New action for the Hilbert-Einstein equations}
\author{L.~D.~Faddeev}
\address{St.Petersburg Department of Steklov Mathematical Institute}

\begin{abstract}
The Hilbert-Einstein equations are derived in the formalism employing the
    imbedding of the space-time into linear 10-dimensional space.
    An extra antisymmetric tensor field is needed for this task.
\end{abstract}

\maketitle

    Einstein's Theory of Gravitation is the most ingenious achievement in the
Theoretical Physics. It has dramatic history, experimental confirmations and
beautiful geometric formulation. Its hamiltonian formulation (see e.~g.
\cite{Faddeev}
and refs. to fundamental papers of Dirac, ADM and others) allows for the
formal quantization. However it is this aspect of the theory which is still
considered unsatisfactory due to the perturbative nonrenormalizability.
Thus the alternative proposals are periodically developed, most prominent of
which are connected with the String Theory.
In this note I discuss one more possibility to modify the mathematical
formulation of the theory retaining the main Hilbert-Einstein equations.

    First I shall remind the basics of the embedding approach to the
    description of metric on space-time, which was discussed by many people
\cite{RT}--\cite{PF}.
    Then I make a trick by extending the set of degrees of freedom by
    promoting the one-forms to full covariant vector fields.
    As a result there will be no extra derivative in the new variational
    equation of motion, whih is the drawback of the Redge-Teitelboim
    equations.

    I begin with the parametrization of the 4-dimensional space-time
$ M^{4} $
    by imbedding into 10-dimensional Euclidian space
$ \RR^{10} $
\begin{equation*}
    f^{A} = f^{A}(x^{\mu}) ,
\end{equation*}
    where
$ f^{A}, A=1,\ldots 10 $ and
$ x^{\mu}, \mu=1,2,3,4 $
    are coordinates in
$ \RR^{10} $ and $ M^{4} $,
    correspondingly.
    For simplicity I shall use the Euclidean signature on
$ M^{4} $,
    which can be easily changed into the Lorentzian one; so the metric on
$ M^{4} $
    and Christoffel's symbols are given by
\begin{equation*}
    g_{\mu\nu} = \partial_{\mu} f^{A} \partial_{\nu} f^{A}
\end{equation*}
    and
\begin{equation*}
    \Gamma_{\lambda,\mu\nu} = \partial_{\lambda}f^{A}
	\partial_{\mu}\partial_{\nu} f^{A}
\end{equation*}
    (see e.~g. V.~Fock monograph \cite{Fock}).
    Let us use these formulas to express the curvature tensor. For this
    I employ the formula
\begin{multline*}
    R_{\mu\nu,\alpha\beta} = \frac{1}{2} \bigl(
	\partial_{\mu}\partial_{\beta} g_{\nu\alpha} +
	\partial_{\nu}\partial_{\alpha} g_{\mu\beta} -
	\partial_{\mu}\partial_{\alpha} g_{\nu\beta} -
	\partial_{\nu}\partial_{\beta} g_{\mu\alpha} - \\
    - g^{\lambda\sigma} (\Gamma_{\lambda,\mu\alpha}\Gamma_{\sigma,\nu\beta} -
    \Gamma_{\lambda,\mu\beta}\Gamma_{\sigma,\nu\alpha})\bigr),
\end{multline*}
    also presented in
\cite{Fock}.

    Substitution the expressions for
$ g_{\mu\nu} $ and
$ \Gamma_{\lambda,\mu\nu} $
    via derivatives of
$ f^{A} $
    gives
\begin{equation*}
    R_{\mu\nu,\alpha\beta} = \Pi^{AB} (f_{\mu\alpha}^{A} f_{\nu\beta}^{B}
	- f_{\nu\alpha}^{A} f_{\mu\beta}^{B}) .
\end{equation*}
    Here I use the short notations
\begin{equation*}
    f_{\mu\alpha}^{A} = \partial_{\mu}\partial_{\alpha} f^{A} , \quad
	f_{\mu}^{A} = \partial_{\mu} f^{A}
\end{equation*}
    and by
$ \Pi^{AB} $
    denote the projector
\begin{equation*}
    \Pi^{AB} = \delta^{AB} - g^{\lambda\sigma} f_{\lambda}^{A} f_{\sigma}^{B}
\end{equation*}
    on the subspace, orthogonal to the tangent space to
$ M^{4} $
    at point
$ x^{\mu} $.
\begin{equation*}
    \Pi^{AB} f_{\mu}^{B} = 0, \quad \Pi^{AB} f^{\mu,B} = 0,
\end{equation*}
    where 
$ f^{\mu,A} = g^{\mu\nu} f_{\nu}^{A} $.

    The expression for
$ R_{\mu\nu,\alpha\beta} $
    clearly is compatible with the symmetry properties of the curvature
    tensor. Let us stress two features of this formula.

    1. Naively we could expect, that
$ R_{\mu\nu,\alpha\beta} $
    is linear in the third derivatives of
$ f^{A} $,
    being linear in the second derivatives of
$ g_{\mu\nu} $.
    However the third derivatives cancel and
$ R_{\mu\nu,\alpha\beta} $
    is a quadratic form of the second derivatives of
$ f^{A} $.

    2. Expression for
$ R_{\mu\nu,\alpha\beta} $
    is covariant in spite of the fact, that it contains only ordinary
    derivatives. Indeed the infinitesimal coordinate variation
\begin{equation*}
    \delta f^{A} = - \epsilon^{\lambda} \partial_{\lambda} f^{A}
\end{equation*}
    induces transformations
\begin{equation*}
    \delta f_{\mu}^{A} = - \partial_{\mu} \epsilon^{\lambda} 
	\partial_{\lambda} f^{A} - \epsilon^{\lambda}
	    \partial_{\lambda} f_{\mu}^{A} ,
\end{equation*}
    corresponding to that of the covariant vector field and
\begin{equation*}
    \delta f_{\mu\alpha}^{A} =
	-\partial_{\mu} \epsilon^{\lambda} f_{\lambda\alpha}^{A}
	-\partial_{\alpha} \epsilon^{\lambda} f_{\mu\lambda}^{A}
	-\epsilon^{\lambda} \partial_{\lambda} f_{\mu\alpha}^{A}
	-\partial_{\mu}\partial_{\alpha} \epsilon^{\lambda} f_{\lambda}^{A} .
\end{equation*}
    The first three terms here correspond to the transformation of the
    covariant tensor field and so are satisfactory. The last unwanted
    term is proportional to the linear combination of vectors
$ f_{\lambda}^{A} $ and is annihilated by projector
$ \Pi^{AB} $.

    The contracted tensor
$ R_{\mu\alpha} = g^{\nu\beta} R_{\mu\nu,\alpha\beta}  $
    and scalar curvature 
$ R $
    can be written via derivatives of the contravariant vector field
\begin{equation*}
    f^{\mu A} = g^{\mu\nu} f_{\nu}^{A}
\end{equation*}
    as follows
\begin{align*}
    R_{\mu\alpha} & = \Pi^{AB} (
	\partial_{\alpha} f_{\mu}^{A} \partial_{\beta}f^{\beta B}
	- \partial_{\beta} f_{\mu}^{A} \partial_{\alpha}f^{\beta B}) \\
    R & = \Pi^{AB} (
	\partial_{\alpha} f^{\alpha A} \partial_{\beta}f^{\beta B}
	- \partial_{\beta} f^{\alpha A} \partial_{\alpha}f^{\beta B})
\end{align*}
    Indeed
\begin{equation*}
    \partial_{\alpha} f^{\mu A} = g^{\mu\nu} \partial_{\alpha} f_{\nu}^{A}
	+ \partial_{\alpha} g^{\mu\nu} f_{\nu}^{A}
\end{equation*}
    and the second term is annihilated by projector.

    With these formulas we prepared our main trick. Let us take
$ f_{\mu}^{A} $
    as generic covariant vector field, put
\begin{equation*}
    g_{\mu\nu} = f_{\mu}^{A} f_{\nu}^{A}, \quad 
    f^{\mu A} = g^{\mu\nu} f_{\nu}^{A}, \quad
	g^{\mu\nu} = f^{\mu A} f^{\nu A}
\end{equation*}
    and take
\begin{equation*}
    \LL = \sqrt{g} R
\end{equation*}
    as lagrangian. It is quadratic in the first derivatives of contravariant
    vector field
$ f^{\mu A} $.
    Then we calculate the variational equations and supplement them by extra
    equations of motion
\begin{equation*}
    \partial_{\mu} f_{\nu}^{A} - \partial_{\nu} f_{\mu}^{A} = 0
\end{equation*}
    with solution
\begin{equation*}
    f_{\mu}^{A} = \partial_{\mu} f^{A},
\end{equation*}
    which will be produced by an additional lagrangian of BF type
\begin{equation*}
    \LL_{1} = B^{\mu\nu,A} (\partial_{\mu} f_{\nu}^{A}
	- \partial_{\nu}f_{\mu}^{A}) ,
\end{equation*}
    where
$ B^{\mu\nu,A} $
    is a set of antisymmetric contravariant tensor densities. Superficially
    this trick has nothing to do with Hilbert-Einstein equations. However
    to my own surprise the variation of
$ \LL $
    will contain
$ R_{\alpha\mu} $.
    More exactly we have formula
\begin{equation*}
    \delta \int \LL d^{4}x = \delta f^{\alpha A} \Gamma_{\alpha}^{A} ,
\end{equation*}
    where
\begin{multline*}
    \Gamma_{\alpha}^{A} = 2\sqrt{g} \bigl(
	-\Pi^{AB}f^{\mu,C}+\Pi^{BC}f^{\mu,A} + \Pi^{AC}f^{\mu,B}
	\bigr)T_{\mu\alpha}^{CB} - \\
    -\sqrt{g} g_{\alpha\mu} \bigl(
	\Pi^{AB}f^{\mu,C}+\Pi^{BC}f^{\mu,A} + \Pi^{AC}f^{\mu,B}\bigr) T^{BC} .
\end{multline*}
    Here
$ T_{\mu\alpha}^{AB} $ and
$ T^{AB} $
    are quadratic forms of the first derivatives, entering the expressions for
$ R_{\mu\alpha} $ and $ R $
\begin{align*}
    T_{\mu\alpha}^{AB} & =
	\partial_{\alpha} f_{\mu}^{A} \partial_{\beta} f^{\beta B} -
	\partial_{\beta} f_{\mu}^{A} \partial_{\alpha} f^{\beta B} , \\
    T^{AB} & = 
	\partial_{\alpha} f^{\alpha A} \partial_{\beta} f^{\beta B} -
	\partial_{\beta} f^{\alpha A} \partial_{\alpha} f^{\beta B} .
\end{align*}
    Furthermore, the variation of
$ \int \LL_{1} d^{4}x $
    gives expression
$ \delta f^{\alpha A} \Sigma_{\alpha}^{A} $,
    where
\begin{equation*}
    \Sigma_{\alpha}^{A} = \Pi^{AB} \partial_{\mu} B^{\mu\nu,B} g_{\alpha\nu}
	- f_{\nu}^{A} f_{\alpha}^{B} \partial_{\mu} B^{\mu\nu,B} .
\end{equation*}
    Altogether the full set of equations of motion is
\begin{equation*}
    \left\{
    \begin{matrix}
	\Gamma_{\alpha}^{A} + \Sigma_{\alpha}^{A} = 0 ,\\
	\partial_{\mu} f_{\nu}^{A} - \partial_{\nu}f_{\mu}^{A} = 0
    \end{matrix}
    \right. .
\end{equation*}
    Projecting the first line on
$ f_{\mu}^{A} $
    we get equations
\begin{gather*}
    R_{\alpha\mu} - \frac{1}{2} g_{\alpha\mu}R + T_{\alpha\mu} = 0 \\
    T_{\alpha\mu} = \frac{1}{\sqrt{g}} g_{\mu\nu} \partial_{\sigma}
	B^{\sigma\nu,B} f_{\alpha}^{B} ,
\end{gather*}
    which together with the second line is equivalent to Hilbert-Einstein
    equations with energy-momentum tensor
    produced by 
$ B $
    field.
    
%    The function
%$ \Lambda $
%    indeed is independent of
%$ x $
%    due to the constraint equation
%\begin{equation*}
%    \nabla^{\mu} (R_{\mu\alpha} - \frac{1}{2}g_{\mu\alpha} R) = 0.
%\end{equation*}
%    The set of equations, orthogonal to
%$ f_{\mu}^{A} $
%    will produce the equations of motion for the
%$ B $
%    field.

    S.~A.~Paston
\cite{Paston}
    has shown that the vertical contribution to
$ \Gamma_{\alpha}^{A} $
    vanishes after the second set of equations of motion is taken into
    account. Thus we have besides the Hilbert-Einstein equations the condition
\begin{equation*}
    \Pi^{AB} \partial_{\mu} B^{\mu\nu,B} = 0 
\end{equation*}
    and so only the horisontal part of 
$ \partial_{\mu} B^{\mu\nu,B} $
    remains unfixed.

    Introduction of the
$ B $-field
    is the price for the success of my trick.
    One must return to this problem of interpretation after more
    close inspection of all equations.

    We finish this note by deriving the variational equations. Due to the
    form of
$ R $
    it is natural to make the variations via contravariant vector field
$ f^{\mu,A} $.
    However the projector
$ \Pi^{AB} $
    contains also covariant fields
\begin{equation*}
    \Pi^{AB} = \delta^{AB} - f_{\mu}^{A} f^{\mu B} .
\end{equation*}
    So the first thing is to find its variation. Using
\begin{equation*}
    f_{\mu}^{A} = g_{\mu\nu} f^{\nu A} ,
\end{equation*}
    we get
\begin{align*}
    \delta f_{\mu}^{A}
	& = \delta g_{\mu\nu} f^{\nu A} + g_{\mu\nu} \delta f^{\nu A} =\\
	& = - g_{\mu\sigma} \delta g^{\sigma\rho} g_{\rho\nu} f^{\nu A}
	    + g_{\mu\nu} \delta f^{\nu A} = \\
	& = -g_{\mu\sigma} (\delta f^{\sigma C}f^{\rho C}
	    + f^{\sigma C} \delta f^{\rho C}) g_{\rho\nu} f^{\nu A}
	    + g_{\mu\nu} \delta f^{\nu A} = \\
	& = \delta f^{\sigma C} (g_{\mu\sigma}\delta^{AC} 
	- g_{\mu\sigma} f^{\rho C} f_{\rho}^{A} - f_{\mu}^{C} f_{\sigma}^{A})
	    = \\
	& = \delta f^{\sigma C}
	    (g_{\mu\sigma} \Pi^{AC}-f_{\mu}^{C}f_{\sigma}^{A}) ,
\end{align*}
    from which follows an elegant formula
\begin{equation*}
    \delta \Pi^{AB}=-\delta f_{\mu}^{A} f^{\mu B} -f_{\mu}^{A} \delta f^{\mu B}
	= -\delta f^{\sigma C}
	    (\Pi^{AC}f_{\sigma}^{B}+\Pi^{CB}f_{\sigma}^{A}).
\end{equation*}

    The derivative of
$ \Pi^{AB} $
    we prefer to express via the derivative of the covariant field.
    The similar calculation gives
\begin{equation*}
    \partial_{\alpha} \Pi^{AB} = -\partial_{\alpha} f_{\sigma}^{C}
	(\Pi^{AC}f^{\sigma B} +\Pi^{CB}f^{\sigma A}) .
\end{equation*}

    Finally for the variation and derivative of
$ \sqrt{g} $
    we have
\begin{equation*}
    \delta\sqrt{g} = -\sqrt{g} \delta f^{\sigma A}f_{\sigma}^{A}, \quad
	\partial_{\alpha} \sqrt{g} = \sqrt{g} f^{\sigma A} \partial_{\alpha}
	    f_{\sigma}^{A}.
\end{equation*}
    Now we have
\begin{multline*}
    \Gamma_{\alpha}^{A} =
	-2\partial_{\alpha}(\sqrt{g} \Pi^{AB} \partial_{\beta} f^{\beta B})
	+ 2\partial_{\beta} (\sqrt{g}\Pi^{AB}\partial_{\alpha}f^{\beta B}) -\\
    -\sqrt{g}(\Pi^{AB} f_{\alpha}^{C} +\Pi^{AC}f_{\alpha}^{B}) T^{BC}
	-\sqrt{g} f_{\alpha}^{A} R
\end{multline*}
    and after differentiating we get the main formula for
$ \Gamma_{\alpha}^{A} $.
    It is instructive to observe that the second derivatives cancel.

    The variation of
$ \LL_{1} $
    gives after using the expression of
$ \delta f_{\mu}^{A} $ via
$ \delta f^{\mu,A} $
\begin{equation*}
    \delta \LL_{1} = -2\partial_{\mu} B^{\mu\nu,A} \delta f_{\nu}^{A}
	= -2\partial_{\mu} B^{\mu\nu,C} (\Pi^{AC}g_{\nu\alpha} -
	f_{\nu}^{A}f_{\alpha}^{C}) \delta f^{\alpha,A} ,
\end{equation*}
    so that
\begin{equation*}
    \Sigma_{\alpha}^{A} = -2 \Pi^{AC}g_{\nu\alpha}\partial_{\mu} B^{\mu\nu,C}
	+2 \partial_{\mu} B^{\mu\nu,C} f_{\alpha}^{C} f_{\nu}^{A}
\end{equation*}
    and again we have explicit separation along tangent vectors and orthogonal
    subspace.

    With this derivation I finish this note. Lot of work for the
    interpretation of the presented trick is ahead. The first thing to do
    is to compare the hamiltonian formulation, following from our lagrangian
    with that of Dirac and ADM.

    I am greatful to S.~A.~Paston for his critical reading of the first draft
    of this paper and to S.~Deser and I.~A.~Bandos
    for valuable comments.


\begin{thebibliography}{0}
\bibitem{Faddeev}
    L.~D.~Faddeev, The energy problem in Einstein's theory of gravitation,
	Sov. Phys. Usp. {\bf 25} (1982) 130--142.

\bibitem{RT}
    T.~Redge, C.~Teitelboim, {\it General relativity a la string: a progress
report}. In Proceedings of the First Marcel Grossmann Meeting, Trieste,
    Italy, 1975. Ed. R.~Ruffini, North Holland, Amstrdam, 1977. P.77.

\bibitem{DPR}
    S.~Deser, F.~A.~E.~Pirani, D.~C.~Robinson, New embedding model of general
    relativity. Phys. Rev. D {\bf 14} (1976) 3301--3303.

\bibitem{PF}
  S.~A.~Paston and V.~A.~Franke,
  ``Canonical formulation of the embedded theory of gravity equivalent to
  Einstein's General Relativity,''
  Theor.\ Math.\ Phys.\  {\bf 153} (2007) 1581
  [Teor.\ Mat.\ Fiz.\  {\bf 153} (2007) 271]
  [arXiv:0711.0576 [gr-qc]].

\bibitem{Fock}
    V.~A.~Fock, {\it Theory of Space, Time and Gravitation}, Moscow, 1961
    (in Russian).

\bibitem{Paston}
    S.~A.~Paston, private communication.

\end{thebibliography}
\end{document}